\documentclass{article}

\newcommand{\EM}{\ensuremath{\eta(548)}}
\newcommand{\OM}{\ensuremath{\omega(782)}}
\newcommand{\AM}{\ensuremath{a(980)}}
\newcommand{\FM}{\ensuremath{f(980)}}
\newcommand{\RM}{\ensuremath{\rho(770)}}

\newcommand{\DR}{\ensuremath{\Delta(1232)}}
\newcommand{\LB}{\ensuremath{\Lambda(1115)}}
\newcommand{\SBa}{\ensuremath{\Sigma^0(1192)}}
\newcommand{\SBb}{\ensuremath{\Sigma^0(1385)}}

\newcommand{\Tlepton}{$\tau$-lepton }

\newcommand{\pion}{\ensuremath{{\pi^0}}}
\newcommand{\kaon}{\ensuremath{{K^0}}}

\newcommand*{\meson}[1]{\ensuremath{#1^\pm}}
\newcommand*{\quarkpair}[1]{\ensuremath{#1\bar #1}}
\newcommand*{\quark}[1]{\ensuremath{#1}}
\newcommand*{\Quark}[1]{\ensuremath{#1}-quark}
\newcommand*{\GeV}[1]{\ensuremath{#1\hbox{\hskip5pt GeV}}}
\newcommand*{\MeV}[1]{\ensuremath{#1\hbox{\hskip5pt MeV}}}

\newcommand*{\ApproxP}[1]{\ensuremath{\approx #1 \%}}

\begin{document}

\begin{center}
{\Large Harmonic quarks: properties and some applications}\par\bigskip
{\large Oleg~A.~Teplov}\par\smallskip
Institute of Metallurgy and Materials Science of RAS, \\ Leninski prospekt 49, Moscow, 119991, Russia.\par
e-mail: teplov@ultra.imet.ac.ru
\end{center}


\begin{abstract}
In this work the investigation of hadronic structures with the help 
of the harmonic quarks is prolonged. The harmonic quark model 
is good at describing the meson structures and the baryon excitations to 
resonances, in particular delta(1232). Harmonic quark reactions 
form the structure of the baryon resonances. Presumed quark structures 
of the mesons eta(548), omega(772), a(980) and f(980) are given. 
It became clear that the some hadronic structures contain the filled 
quark shells. The kinetic quark energy in the basic charged mesons 
are enough small for a using of perturbative methods. The following 
topics are briefly considered and discussed: harmonic quark series 
and its boundaries, the d-quark peculiarity, parallel quark series 
and quark mixing. The boundaries of quark chain can are closely 
related to a weak interaction. The cause of the quark mixing is 
probably an existence of the parallel quark chain and the special 
properties of the \Quark{d} in the main quark chain. The new mass 
equation is found. It is probably a manifestation of Higgs
mechanism. Using this equality enables to improve the accuracy of 
the harmonic quark masses calculation to 0.005\%. The strong 
interaction should take into account the harmonic quark annihilation.   

\end{abstract}


\section{Introduction}

In article~\cite{my1} on the basis of the harmonic quark oscillator we built 
a simple quark model, in which the quark masses are connected by the 
recurrent equations:  

\begin{equation}\label{MQ}
   {m_n\over m_{n-1}}= {\pi\over(4-\pi)} = 3.6597924 \stackrel{\rm def}{=} MQ
\end{equation}
or so
\begin{equation}\label{qmass}
  m_n = m_0\left[{\pi\over4-\pi}\right]^n
\end{equation}

where $m_0$ is a hypothetical initial quark mass; $m_n$ is the quark mass of 
quark flavor number $n$.\\
Further we calculated the harmonic quark masses with the 
\ApproxP{0.03} inaccuracy (table 1). 

\begin{center}
 Table 1. The masses of harmonic quarks. 
  \medskip
  \begin{tabular}{|l|c|c|c|c|c|c|c|}
    \hline
   n (flavor) & 1 & 2 & 3 & 4 & 5 & 6 & 7\\
    \hline
    quark & $d$ & $u$ & $s$ & $c$ & $b$ & $t$ & $b'$\\
    \hline
    mass (MeV)  & 28.815 & 105.456 & 385.95 & 1412.5 & 5169.4 & 18919 & 69239\\ 
    \hline
  \end{tabular}
\end{center}
Then we started to show actual serviceability of the harmonic quarks. 
So, to apparent results concern a solution 
of secret of an origin of mass of a muon and interpretation quark 
structure of a neutral pion. It seems that \textbf{neutral pion with
great share of probability is a stationary harmonic oscillator based on
quark-antiquark \textit{u}-pair}. The pion mass is more the oscillator mass on
0.7 MeV, that together with Coulomb energy of pair provide a kinetic energy of
 \Quark{u}s up to some MeVs.
 The next obvious conclusion is that 
\textbf{muon is a successful attempt of Nature to explicitly fix the 
single \Quark{u} mass state as a lepton suppressing color and 
fractional charge.} So, main decay canals of \meson{\pi} and \meson{K} is
$\mu$ and $\nu$, then hence \Quark{d} and \Quark{s} disintegrate in these
 mesons much more often than \Quark{u}.
Probably electric and colour charges of \Quark{d} and \Quark{s} transfer to 
\Quark{u} and then $\mu$ shall born. We do not know, as a color charge is 
suppressed and as the lepton numbers are formed. Important only that such 
process can occur but it is beyond the framework of the present article.\\ In 
this work we shall prolong an examination of hadronic structure 
with help of the harmonic quarks and quark oscillators.
Besides we shall prolong to study the properties of our harmonic model and 
we shall discuss a controversy on problems and corollaries generated 
by this model.

\section{Harmonic quarks in hadrons}
The harmonic quark applications to hadronic structures are below started. 

\subsection{Quark reactions}

Harmonic quarks demonstrate a very good serviceability at the description
of baryons and their resonances. We shall consider a simple quark 
reactions only for the following pairs of particles: \LB\ and \SBa, 
\LB\ and \SBb, nucleon and \DR.

\begin{enumerate}
\item
  After the electromagnetic transition of \SBa\ to \LB\ the baryon mass
  decreases by \MeV{76.96}.  The transition of \quarkpair{u} quark
  pair to the harmonic oscillator \\ $u$ + $\bar u$ = ($\quarkpair{u})_{ho}$ \\ is 
  accompanied by the mass decrease of \MeV{76.63}.
\item
  Now we shall write a simple quark reaction of \LB\ excitation up to \SBb\
  with the mass of \MeV{1383.7\pm1.0}~\cite{PDG}:
  $$
  \SBb=\LB+\quarkpair{u}+\quarkpair{d}
  $$
  From here the mass of \SBb\ is 1115.7 + 2(105.4 + 28.8) = \MeV{1384.1}, 
  which is in complete accord with the experimental measurements. 
  The main channel of the \SBb\ triplet decay (88\%) is \LB$\pi$.
\item
  The pion-nucleon scattering has the first peak at the aggregate
  energy of \MeV{1232}, which is known as the first delta baryon.  Note how
  simply and gracefully this resonance can be obtained from the
  transformation reaction of a $d$-quark pair into a $u$-quark pair
  $$ d_{\rm nucleon}+{\bar d}_{\rm pion}=u_{\rm delta}+{\bar u}_{\rm delta} $$
  with the mass of \DR\ equal to 
  $939.6 + 139.6 + 2(105.4 - 28.8) = \MeV{1232.4}$,
  this is also in a complete accord with experimental data.
\end{enumerate}

\subsection{Quark structure of some light mesons}
We shall now determine the plausible structure of light mesons by
``sequentially filling'' them using quark masses obtained.  That is, we
would first ``fill'' the meson mass with the heaviest possible quark,
then again ``fill'' the rest of its mass with the heaviest possible
quark, etc.  As pion and kaon were already discussed, let's take the
next particle: \EM.  The sequential filling of its mass immediately
produces the harmonic quark structure \quark{s+u+d+d} with the desired mass
of \MeV{549}.  Quite acceptable.  It's rather easy to interpret
this result with the help of a harmonic oscillator concept.  The total
mass of the \quark{su} pair is equal to the mass of stationary harmonic
oscillator based on a strange quark-antiquark pair.  The remaining light
quark pair forms a filled outer neutral shell of $1s^2$ kind.  On the
whole the \EM\ structure is similar to helium atom---a heavy
nucleus formed by a stationary harmonic oscillator and a relatively light
filled outer shell.  The analogy can also be drawn to helium nuclei,
which has two filled $1s^2$ shells of two protons and two neutrons.
Beside that, note the similarity between \EM\ and \pion.  Both particles
supposedly contain a harmonic oscillator.

Now let's consider the structure of \OM\ vector meson, which is
generally agreed to contain hidden strangeness.  The strange
\quarkpair{s} pair has total mass of \MeV{772} which is only \MeV{10}
less then the meson mass. The \OM\ differs from all particles formerly
discussed, as it contains the separated quark-antiquark pair of the same
flavor. This pair is completely separated to distinct entities with certain
quark-related restrictions.  As in \meson{\pi} and \meson{K} mesons case
the \quark{s}-quarks of \OM\ demand an additional energy: a kinetic energy and
probably the energy to hold the quark spins in the same direction.

We will not discuss the \RM, $K^\star$ and $\eta(958)$ mesons, for their
structure is somewhat more complicated, but will say a few words about
\AM\ with mass value of \MeV{984.8\pm1.4} and \FM: \MeV{980\pm10}.  The
sequential filling provides the \quark{s+s+u+u} quark structure with total
mass value of \MeV{982.9}.  Another reasonable result.  Though this case
is more intricate as the \AM\ is a triplet.  Therefore this structure
supposedly describes the \FM\ meson only.  There's two filled neutral
shells of a $1s^2$ kind here based on \quarkpair{s} and \quarkpair{u}.
The \AM\ triplet structure resembles the structure of \RM triplet, as they
probably differ by one quark-antiquark \quarkpair{u}-pair only
(\MeV{211}) with the extra kinetic energy of several MeVs.

Thus we can see that harmonic quarks provide us with a completely new
insight of meson structures. These examples of the harmonic quark model 
usefulness could be easily prolonged. Though the ones presented here 
should suffice to realize the model potential. It seems we had actually 
found a new powerful tool of hadronic (and other particles) structure 
determination.

\subsection{Quark state in \meson{\pi}, \meson{K} and \meson{b} mesons}

Using precise values of quark masses one can easily obtain quark velocity
limits and minimal inter-quark distances in charged pseudo-scalar mesons
of \meson{\pi} kind.  Estimating minimal inter-quark distances we would
assume that a colur interaction on these distances is small, hence
the quark kinetic energy would transform to Coulomb energy as the quarks
draw near each other.  The maximum velocities presented in table~2 are 
given in light unit and calculated at the force balance moment between
Coulomb and color fields.

\begin{center}
  Table 2. Maximum quark speeds and minimal inter-quark distances\par
  in \meson{\pi}, \meson{K} and \meson{b} mesons
  \par\medskip
  
  \begin{tabular}{|c|c|c|c|c|}
    \hline
          & Additional       & maximum   & maximum   & minimal     \\
    Meson & quark energy     & $u$-quark & 2nd quark & inter-quark \\
          & MeV              & velocity  & velocity  & distance,fm \\
    \hline
    \meson{\pi} ($ud$) & 5.32 & 0.146 & 0.487 & 0.052\\
    \meson{K} ($us$)   & 2.32 & 0.183 & 0.051 & 0.12 \\
    \meson{b} ($ub$)   & 4.9$\pm$0.5 & 0.29 & 0.006 & 0.057\\
    \hline
  \end{tabular}
\end{center}

It can be concluded that quarks in charged mesons are not very
relativistic.  It's the only lightest \Quark{d} that reaches half light
speed in a pion. The data in table~2 convince us that a pertubative methods
would be used for investigation of these mesons on the based of the 
harmonic quarks. 

\section{Harmonic quark properties and discussion}

We shall now investigate the consequences of the presented harmonic 
quark family and other matters ensuing from the model and the quark
masses.  It's a well-known fact that harmonic oscillators play a
significant part in classic field theory~\cite{peskin,manin}, are a
``perfect model'' and ``everywhere''~\cite{manin}.  The model quark spectrum
 bears a resemblance to the simple harmonic oscillator (with
$(n+{1\over2})\omega$ energy spectrum).  Both of them are equidistant
with the only difference being that the first on a normal and the second on
a logarithmic scale~(\ref{MQ}).  Just as a simple harmonic oscillator 
has the zero state, we can single out the zero state namely the zero quark 
(eq. (\ref{qmass}) at $n=0$), that is the hypothetical initial mass (energy) 
$m_0\cong\MeV{7.87}$.  The author believes that harmonic quark masses are 
actually their physical masses. Anyway, there are reasons to believe 
that this harmonic quark energy spectrum is the spectrum of the 
eigenstates of a certain interacting field. If that is the case, 
we are up to solving the inverse problem: to find the field using its 
eigenstates.
The ground state of the quark-antiquark pair is a harmonic oscillator
based on it.  It is probably the most coupled state, for all two-quark 
meson masses are greater than the masses of corresponding 
harmonic oscillators:  pions masses are greater than \quark{u}-oscillator
energy, the same with kaons and \quark{s}-oscillator, etc.
If the coupled energy were any greater, the quarks should probably lose
their individuality. 

\subsection{Quark family boundaries}

As the quark model presented forms a rigid series from the quark family,
it's inevitable to wonder about the lower and upper boundaries of
it.  There is not much doubt about the lightest quark of the series---it's 
the \Quark{d} (\MeV{28.8}).  The particle physics 
doesn't currently know the charged meson lighter than \meson{\pi}, which
contains \Quark{d} and \quark{u}.  Quark lighter than \Quark{d}
would have \MeV{7.87} mass value, and corresponding meson \MeV{$40-50$}.
The termination of the series from below may be occurs because of the
electron-positron field.  The upper boundary of the quark series, i.e.\ the
heaviest existing quark is still a mystery.  For heavy quarks \quark{t}, 
\quark{b'} and \quark{t'} the model mass values are~\GeV{19}, \GeV{69} 
and \GeV{253} respectively.

The \Quark{t'} mass would be three times more that of $W$-boson.
Therefore it is only logical to assume that the termination reason from
above could be weak interaction with its characteristic distance and energy
scales.

Unifying the reasons one can say that the quark series may be limited from
both sides by electroweak interaction.  Thus in the energy range from
zero to $W$ and $Z$ bosons there are seven harmonic quark flavors.

It's interesting to mention that a hypothetical initial quark mass
$m_0=\MeV{7.9}$ in~(\ref{MQ}) lies very close to the \Quark{d} bare mass
and the mass value of ${1\over2}(u+d)$ combination, computed with the help
of  $\overline{MS}$-schemes and lattice simulations~\cite{PDG}.

\subsection{Parallel quark series}
Alongside with the quark series already discussed, which we shall name
``basic series'', one can easily imagine another ``unrealized'' parallel
quark series with the same harmonic relations~(\ref{qmass}) and the same
mass values.  The only difference between the series is the ``charge
shift'' by one position.  Thus, parallel quarks with mass values of
\MeV{28.8} and \MeV{105.4} would have the charges of $\pm{2\over3}$ and
$\pm{1\over3}$ respectively, etc.  The reasons for Nature preference of
the basic series to the parallel series are unknown at present.  It may be
somehow connected with the special features of the \Quark{d}.  For
example, there could be some reasons for the lightest quark to
necessarily have the charge of $\pm{1\over3}$.  The \Quark{d} position 
in the quark chain is distinguished already.  Any other quark has two 
neighbors, which can be thought of as an upward and downward excitation, 
but the \Quark{d} has only an upward excitation.  In a sense it is an 
``inferior'' quark. It can be to some extent considered as an appendix 
to the \Quark{u}. The isotopic properties of the \quark{d}- and \Quark{u}s 
is likely to lie there.  Furthermore this could be an approach to 
solving the quark mixing problem, as the mixing can be a partial 
manifestation of the parallel series.

\subsection{Leptons}

The successful decision of a muon mass problem on the basis of \Quark{u}
stimulate us to review the process of the \Tlepton mass formation too.  
The \Quark{c} mass is apparently close to the \Tlepton mass and the author
believes that the \Tlepton may be formed on the mass basis of \Quark{c}.  
The existing mass difference between them of \MeV{364} should have 
a good reason and needs to be explained.  This is less than the 
neighboring \Quark{s} mass and moreover it can't be the 
\Quark{s}. Otherwise we shall receive a meson. At the same time \Tlepton 
mass is less than the mass of the harmonic c-oscillator and also than 
the lightest $D$-meson mass, which contains explicit charm.  One can 
suppose that this additional energy is required for suppression both color 
and fractional electrical charges on the \Quark{c}, but it is not enough 
for $D$-meson formation. Both the \Tlepton and the muon in it are similar.   
Their masses are located below masses of the first mesons of the 
appropriate quarks. One could also suppose that the massive
leptons could only be based on harmonic quarks with the $\pm{2\over3}$
charge, ignoring the quarks with charge of $\pm{1\over3}$.  The existence of
the super-heavy charged lepton based on the harmonic \Quark{t} is not
likely due the following: there are no known hadrons with the explicit
\quark{t}-flavor, which means that solitary \Quark{t} existence without
\quark{t}-antiquark is suppressed for some reason, therefore the
individual mass basis to form the super-heavy lepton on does not exist.
Furthermore, the experimental and theoretical works testify to that
there are only three light neutrinos~\cite{PDG}, which is also a good
argument against the existence of the fourth pair of leptons.

From the supposed point of view there is necessary the some updating
the symmetry with lepton-quark generations.

\begin{itemize}
\item The harmonic quarks are bound in one chain, instead of broken-down 
on three generations.
\item The muon and the \Tlepton are formed on the mass basis of quarks 
with a charge ± 2/3, that does the quarks as though by more fundamental 
fermions. 
\item An electron is not bound in any way with quark set 
and it, probably, really formed on the basis of QED vacuum. 
\end{itemize}

\subsection{Is it the Higgs mechanism manifestation?}

The muon formation of \Quark{u} mass state singles out this quark from
the quark series too.  The \Quark{u} is a lightest quark with the charge of
${2\over3}e$ which has both two neighbors.  It is as though it was a first
full-fledged quark with a full-fledged properties and should be
considered in details as such.  The fully color neutral group based 
on this quark, i.e.\ three colors and three anti-colors (let's name it
\quark{6u}-boson) turns out to have the mass value of
\MeV{632.736\pm0.18}.  And the total mass of the first two neutral
particles \pion\ and \kaon\ also have the same value:
\MeV{632.649\pm0.03}.  Pion is a truly neutral particle, and
kaon in the long-term sense (mean integral) is also truly neutral, as
it is able to pass between ``particle'' and ``antiparticle'' states.
Thus we have a mass equality of two truly neutral groups.  The author
believes this not to be a coincidence, but the component of Higgs mechanism,
the true meaning of which is still unknown.  Nonetheless we can use this
equality. The inaccuracy of \Quark{u} mass determination can be
improved six fold with its help, and harmonic quark masses can be
obtained with 0.005\% inaccuracy.

\begin{center}
  Table 3. Precise mass values of the harmonic quarks, MeV
  \small
  \begin{tabular}{|c|cccccc|c|}
    \hline
    meson & d & u & s & c & b & t & error,\%\\
    \hline
    \meson{b}~\cite{my1}   &28.815&105.456&385.95&1412.49&5169.4&18919.0&$\pm$0.030\\
    $\pion+\kaon$ &28.8108&105.441&385.894&1412.29&5168.7&18916.3&$\pm$0.005\\
    \hline
  \end{tabular}
\end{center}

\section{Conclusion}
The author believes that harmonic quark masses are actually their 
physical masses. We can see that the harmonic quark annihilation plays 
a significant role in strong interaction. The harmonic quark annihilators 
can be directly bound with a bundle of space and can be appear the 
elementary boson excitations of a vacuum. The harmonic bound states of 
quarks should find the place in the QCD Lagrangian. The simple recurrent 
equation for quark masses and their precise binding to energy scale will 
considerably reduce the number of the free parameters of the Standard Model. 



\begin{thebibliography}{99}

\bibitem{my1}
  O.~A.~Teplov, arXiv:hep-ph/0306215.
\bibitem{PDG}
  K.~Hagiwara {\em et al.} (Particle Data Group), Phys.~Rev.~D.{\bf 66} 
  (2002) 010001.
\bibitem{peskin}
  M.~E.~Peskin and D.~V.~Schroeder, {\em  An introduction to quantum 
  field theory}, Addison-Wesley 1995.
\bibitem{manin}
  I.~Yu.~Kobzarev and Yu.~I.~Manin, {\em Elementary Particles: Mathematics, 
  Physics and Philosophy}, Dordrecht: Kluwer Academic Publishers 1989.

\end{thebibliography}
\end{document}